\documentclass[aps,prb,preprint,showpacs,superscriptaddress,floatfix]{revtex4}
\usepackage{graphics}
\usepackage{epsfig}
\usepackage{amsmath}
\usepackage{amssymb}
\usepackage{calc}
\newcommand{\beq}{\begin{equation}}
\newcommand{\eeq}{\end{equation}}
\newcommand{\bea}{\begin{eqnarray}}
\newcommand{\eea}{\end{eqnarray}}

\begin{document}
\title{On the geometry of four qubit invariants}
\author{P\'eter L\'evay}
\affiliation{Department of Theoretical Physics, Institute of Physics, Budapest University of Technology and Economics, H-1521 Budapest, Hungary}
\date{\today}
\begin{abstract}
The geometry of four-qubit entanglement is investigated. 
We replace some of the polynomial invariants for four-qubits introduced recently by new ones of direct geometrical meaning. 
It is shown that these invariants describe four points, six lines and four planes in complex projective space ${\bf CP}^3$. For the generic entanglement class of stochastic local operations and classical communication they take a very simple form
related to the elementary symmetric polynomials in four complex variables.
Moreover, their magnitudes are entanglement monotones that fit nicely into the
geometric set of $n$-qubit ones related to Grassmannians
of $l$-planes found recently.
We also show that in terms of these invariants the hyperdeterminant of order $24$ in the four-qubit amplitudes takes a more instructive form than the previously published expressions available in the literature.
Finally in order to understand two, three and four-qubit entanglement in geometric terms we propose a unified setting based on ${\bf CP}^3$ furnished with a fixed quadric.
\end{abstract}
\maketitle{}

\section{Introduction}

Recently the problem of characterizing $n$-qubit entanglement classes has generated considerable interest. This problem was raised within the context of quantum information theory regarding the physical phenomenon of entanglement as a resource. In order to exploit the capabilities encoded in this resource for different tasks of quantum information processing we have to somehow measure it.
During the past few years a number of useful entanglement measures for pure states has been appeared \cite{Wootters,Kundu,Wong,Meyer,Luque,Emary,Osterloh,Thibon,Levay}. 
Classifications up to four qubits has been appeared \cite{Dur,Verstraete,Miyake}and the interesting geometric structures associated with entangled pure states has been noted \cite{Miyake,Bengtsson,Brody,Mosseri,Lev1,Lev2,Levay,Heydari}.

In our previous set of papers \cite{Levay,Lev1,Lev2} using some results from twistor theory we initiated an approach
for understanding $n$-qubit entanglement in geometric terms.
We have shown that this problem can be completely solved for three-qubits, and we obtained partial results for a special subclass characterized by $n$-qubit entanglement monotones.
The aim of the present paper is to add some interesting new results 
on the geometry of four-qubit SLOCC (stochastic local operations and classical communication\cite{Bennett}) invariants.
Such invariants has been introduced in [5], where the Hilbert series of the algebra of invariants has been found.
This result enabled the authors to construct a complete set of four algebraically independent invariants of degree $2,4,4,6$ in the complex coefficients characterizing the four-qubit entangled pure state.
The values of these invariants on the SLOCC orbits of Ref. [11]
were also given. Moreover, an explicit formula for the hyperdeterminant of degree 24 was also obtained.
The authors of Ref. [5] has conjectured that some of the invariants might have a geometric meaning.
In this paper we show that this is indeed the case.

In Section II. we present a new set of four invariants, by replacing two from the ones of Ref. [5]. In Section III. we clarify the geometric meaning of these invariants in terms of the geometry of ${\bf CP}^3$ the complex projective space. 
In Section IV. we show that using our new set of invariants the expression for the 24th order hyperdeterminant takes a more instructive form than the one that can be found in [5].
Moreover, it turns out that the entanglement classes invariant under
SLOCC transformations take a very simple
form
related to the elementary symmetric polynomials in four complex variables.
The magnitudes of our invariants turn out to be entanglement monotones that fit nicely into the
geometric set of $n$-qubit entanglement monotones related to Grassmannians
of $l$-planes in ${\bf C}^{L}$ with $L=2^n$, $l\leq L$ found recently.
Finally our conclusions and some comments are left for Section V.

\section{Invariants}

Let us write an arbitrary four qubit state in the form

\beq
\vert\Psi\rangle =\sum_{i,j,k,l=0}^1Z_{ijkl}\vert ijkl\rangle\in {\bf C}^2\otimes{\bf C}^2\otimes{\bf C}^2\otimes{\bf C}^2
\label{4qubit}
\eeq
\noindent
where $\vert ijkl\rangle={\vert i\rangle}_1\otimes{\vert j\rangle}_2\otimes{\vert k\rangle}_3\otimes{\vert l\rangle}_4$.
Following [5] we introduce decimal notation for $Z_{ijkl}\equiv Z_r$
where $r=8i+4j+2k+l$ and the matrices

\beq
{\cal L}=\begin {pmatrix}Z_0&Z_4&Z_8&Z_{12}\\Z_1&Z_5&Z_9&Z_{13}\\Z_2&Z_6&Z_{10}&Z_{14}\\Z_3&Z_7&Z_{11}&Z_{15}\end {pmatrix}\equiv\begin {pmatrix} {\bf A},&{\bf B},&{\bf C},&{\bf D}\end {pmatrix},
\label{L}
\eeq

\beq
{\cal M}=\begin {pmatrix}Z_0&Z_2&Z_8&Z_{10}\\Z_1&Z_3&Z_9&Z_{11}\\Z_4&Z_6&Z_{12}&Z_{14}\\Z_5&Z_7&Z_{13}&Z_{15}\end {pmatrix}\equiv \begin {pmatrix}{\cal A}^T&{\cal C}^T\\{\cal B}^T&{\cal D}^T\end {pmatrix}
\eeq

\beq
{\cal N}=\begin {pmatrix}Z_0&Z_1&Z_8&Z_9\\Z_2&Z_3&Z_{10}&Z_{11}\\Z_4&Z_5&Z_{12}&Z_{13}\\Z_6&Z_7&Z_{14}&Z_{15}\end {pmatrix}\equiv \begin {pmatrix} {\cal A}&{\cal C}\\{\cal B}&{\cal D}\end {pmatrix}.
\eeq
\noindent
Here ${\bf A}, {\bf B}, {\bf C}$, ${\bf D}\in {\bf C}^4$  are considered as
four column vectors and ${\cal A}, {\cal B}, {\cal C}, {\cal D}$ are $2\times 2$ matrices with $T$ referring to transposition.
We wish to describe the geometry of four qubit entanglement in terms of the four vectors $A_{\alpha}, B_{\beta}, C_{\gamma}$ and $D_{\delta}$ living in ${\bf C}^4$  where $\alpha, \beta, \gamma,\delta =0,1,2,3$. Hence we regard the matrix ${\cal L}$ as fundamental.
The matrices ${\cal M}$ and ${\cal N}$ will be used later.

We are interested in studying a subset of polynomials in the complex numbers $Z_r, r=0,\dots 15$ that are invariant under the SLOCC group of stochastic local operations and classical communication i.e. $SL(2, {\bf C})^{\otimes 4}$.
Such transformations are of the form

\beq \vert\Psi\rangle \mapsto (S_1\otimes S_2\otimes S_3\otimes S_4)\vert\Psi\rangle,
\eeq
\noindent
where $S_m\in SL(2, {\bf C}), m=1,2,3,4$, and $\vert\Psi\rangle$ takes the (\ref{4qubit}) form with the indices of $S_m$ referring to the label of the ${\bf C}^2$ in the tensor product they are acting on.

In order to define the SLOCC invariants we wish to propose we will introduce two extra structures on ${\bf C}^4$.
The first one is a bilinear form $g:{\bf C}^4\times {\bf C}^4\to {\bf C}$
such that for two vectors ${\bf A},{\bf B}\in {\bf C}^4$
we have
\beq
\label{bili}
({\bf A}, {\bf B})\mapsto g({\bf A}, {\bf B})\equiv {\bf A}\cdot {\bf B}=
g_{\alpha\beta}A^{\alpha}B^{\beta}=A_{\alpha}B^{\alpha}
\eeq
\noindent
where
\beq
\label{expl}
g=\begin {pmatrix} 0&0&0&1\\0&0&-1&0\\0&-1&0&0\\1&0&0&0\end {pmatrix}=
\begin {pmatrix} 0&1\\-1&0\end {pmatrix}\otimes
\begin {pmatrix} 0&1\\-1&0\end {pmatrix},
\eeq
\noindent
$\alpha,\beta =0,1,2,3$ and summation for repeated indices is understood.
Notice that the matrix of our symmetric bilinear form can be written in a tensor product form $g={\varepsilon}\otimes {\varepsilon}$ where ${\varepsilon}$
is invariant under the group $SL(2, {\bf C})$, i.e. we have $S{\varepsilon}S^T={\varepsilon}$ with $S\in SL(2, {\bf C})$.  
Moreover, since

\beq
A_{\alpha}=Z_{00kl}, \quad B_{\beta}=Z_{01kl}, \quad C_{\gamma}=Z_{10kl}, \quad D_{\delta}=Z_{11kl},\quad \alpha,\beta,\gamma,\delta=0,1,2,3,\quad k,l=0,1
\eeq
\noindent
quantities involving this symmetric bilinear form are automatically invariant
with respect to $SL(2, {\bf C})\otimes SL(2, {\bf C})$ transformations
of the the third and fourth qubit, i.e. the ones of the form $I\otimes I\otimes S_3\otimes S_4$.
Since $SL(2, {\bf C})\otimes SL(2, {\bf C}))/{\bf Z}_2\simeq SO(4, {\bf C})$ 
it follows that greek indices like ${\alpha}, {\beta}$ etc can also be regarded as vector indices under $SO(4, {\bf C})$. This conversion of complex four-vectors into complex $2\times 2$ matrices
has already been used elsewhere to connect the results of twistor theory to the geometry of entanglement \cite{Lev1,Lev2}.
Hence the columns of the matrix ${\cal L}$ of (\ref{L}) transform as vectors under transformations of the form $I\otimes I\otimes S_3\otimes S_4$
and as the $00$, $01$, $10$ and $11$ components of a tensor under the ones with the form $S_1\otimes S_2\otimes I\otimes I$.

The second structure, as we will see, is related to the notion of duality in ${\bf CP}^3$.
For the vectors ${\bf A}, {\bf B}, {\bf C}$ and ${\bf D}$ let us introduce their {\it duals} as 

\beq
a_{\alpha}\equiv -{\epsilon}_{\alpha\beta\gamma\delta}B^{\beta}C^{\gamma}D^{\delta},\quad
b_{\beta}={\epsilon}_{\alpha\beta\gamma\delta}A^{\alpha}C^{\gamma}D^{\delta},\quad
c_{\gamma}\equiv {\epsilon}_{\alpha\beta\gamma\delta}A^{\alpha}B^{\beta}D^{\delta},
\quad
d_{\delta}\equiv -{\epsilon}_{\alpha\beta\gamma\delta}A^{\alpha}B^{\beta}C^{\gamma}.
\eeq
\noindent
Clearly these quantities transform as vectors under transformations of the form
$I\otimes I\otimes S_3\otimes S_4$ and a straightforward calculation shows 
that ${\bf a}, {\bf b}, {\bf c}$ and ${\bf d}$ behave under the ones of the form $S_1\otimes S_2\otimes I\otimes I$ as the $11$, $10$, $01$ and $00$ components of a tensor respectively.

Let us now introduce the notation ${\bf A}\wedge {\bf B}$ (a bivector)
corresponding to the antisymmetric matrix ${A_{\alpha}B_{\beta}-A_{\beta}B_{\alpha}}$ and the one

\beq
L\equiv{\rm Det}{\cal L}={\epsilon}^{\alpha\beta\gamma\delta}A_{\alpha}B_{\beta}C_{\gamma}D_{\delta}.
\eeq

Now the SLOCC invariants we wish to propose are

\beq 
I_1=\frac{1}{2}({\bf A}\cdot{\bf D}-{\bf B}\cdot {\bf C}).
\eeq
\noindent
\beq
I_2=\frac{1}{6}\left(({\bf A}\wedge {\bf B})\cdot({\bf C}\wedge{\bf D})+
({\bf A}\wedge{\bf C})\cdot({\bf B}\wedge{\bf D})-\frac{1}{2}
({\bf A}\wedge{\bf D})^2-\frac{1}{2}({\bf B}\wedge{\bf C})^2)\right)
\eeq
\noindent
\beq
I_3=\frac{1}{2}({\bf a}\cdot{\bf d}-{\bf b}\cdot{\bf c}).
\eeq
\noindent
\beq
I_4=L
\eeq
Here quantities like $({\bf A}\wedge {\bf B})\cdot({\bf C}\wedge{\bf D})$ are defined as

\beq
\label{cdot}
({\bf A}\wedge {\bf B})\cdot({\bf C}\wedge{\bf D})=(A_{\alpha}B_{\beta}-A_{\beta}B_{\alpha})(C^{\alpha}D^{\beta}-C^{\beta}D^{\alpha})=
2(({\bf A}\cdot {\bf C})({\bf B}\cdot{\bf D})-({\bf A}\cdot{\bf D})({\bf B}\cdot{\bf C})).
\eeq

The first of our invariants $I_1$ takes the form

\beq
I_1=\frac{1}{2}H=\frac{1}{2}(Z_0Z_{15}-Z_1Z_{14}-Z_2Z_{13}+Z_3Z_{12}-Z_4Z_{11}+Z_5Z_{10}+Z_6Z_9-Z_7Z_8),
\eeq
\noindent
showing that $I_1=\frac{1}{2}H$ where $H$ is one of the basic invariants of Ref.[5]. It is just a special case of the $n$-tangle with $n$ even introduced earlier by Wong and Christensen \cite{Wong}.
Reverting to binary notation it is easy to show that $I_1$ is also a permutation invariant \cite{Luque,Wong,Levay}.

Our last invariant $ I_4$ is just $L={\rm Det}{\cal L}$ 
an invariant also introduced by the authors of [5].
These authors have also introduced two more invariants of order four and six denoted by $M$ and $D$ respectively
(the first of them being just {\it minus} the determinant of our matrix ${\cal M}$. They have shown after obtaining the Hilbert series that the invariants $H$, $L$, $M$ and $D$ are algebraically independent and complete. 
Here instead of the invariants $M$ and $D$ we prefer the new ones $I_2$ and $I_3$. As we will see the set $(I_1, I_2, I_3, I_4)$ is of geometrical significance. Moreover, it turns out that the values of these invariants on the generic SLOCC orbit \cite{Verstraete} of four-qubit entangled states are just the elementary symmetric polynomials in four complex variables.
We will also show that in terms of this new set of invariants the explicit formula for the hyperdeterminant of degree 24 takes a more instructive form
then the corresponding one of Ref. [5].

\section{The geometric meaning of four qubit invariants}
\subsection{The invariant $I_2$}

Let us now explain the structure of $I_2$! In order to do this
we introduce another ${\bf C}^4$  corresponding to the four-vector structure
also present in the {\it first two} indices of $Z_{ijkl}$.
Converting the first two spinor indices to vector ones labelled by $\mu=0,1,2,3$,
what we obtain is a "vector-valued" four vector $Z_{\mu\alpha}=({\bf A},{\bf B},{\bf C},{\bf D})^T$. (Alternatively regarding $Z_{\mu\alpha}$ as a $4\times 4$ matrix we obtain the matrix ${\cal L}$ of Eq. (2).)
Let us now also supply this new copy of ${\bf C}^4$ with the bilinear form $g$
known from Eq. (\ref{bili}) with matrix $g_{\mu\nu}$ $\mu,\nu =0,1,2,3$.
Define now the second exterior power of a matrix as the map

\beq
{\bigwedge}^2:{\bf C}^{n\times n}\to {\bf C}^{\left({n\atop 2}\right)\times\left({n\atop 2}\right)}
\eeq
\noindent
which takes an $M_{\mu\nu}\in{\bf C}^{n\times n}$, $0\leq \mu,\nu\leq n-1$ to

\beq
M^{(2)}\equiv\left({\bigwedge}^2 M\right)_{IJ}\equiv M_{\mu_1\nu_1}M_{\mu_2\nu_2}-M_{\mu_1\nu_2}M_{\mu_2\nu_1}
\eeq
\noindent
where $I=\{\mu_1,\mu_2\}$ with $0\leq \mu_1<\mu_2\leq n-1$ and 
 $J=\{\nu_1,\nu_2\}$ with $0\leq \nu_1<\nu_2\leq n-1$.
For the $4\times 4$ matrix $g_{\mu\nu}$ of our bilinear form $g$  
we have

\beq                                                                            \label{expl2}                                                                   G_{IJ}\equiv g^{(2)}_{IJ}=g^{(2)IJ}=\begin {pmatrix} 0&0&0&0&0&1\\0&0&0&0&1&0\\0&0&-1&0&0&0\\0&0&0&-1&0&0\\
0&1&0&0&0&0\\1&0&0&0&0&0\end {pmatrix} 
\eeq                                                                            \noindent
where $I,J=01,02,03,12,13,23$

Let us now introduce the "bivector-valued" $4\times 4$ antisymmetrix Pl\"ucker matrix

\beq
\label{Plucker}
P_{\mu\nu}=\begin{pmatrix} 0&{\bf A}\wedge{\bf B}&{\bf A}\wedge{\bf C}&
{\bf A}\wedge{\bf D}\\-{\bf A}\wedge{\bf B}&0&{\bf B}\wedge{\bf C}&{\bf B}\wedge{\bf D}\\-{\bf A}\wedge{\bf C}&-{\bf B}\wedge{\bf C}&0&{\bf C}\wedge{\bf D}\\
-{\bf A}\wedge{\bf D}&-{\bf B}\wedge{\bf D}&-{\bf C}\wedge{\bf D}&0\end{pmatrix}
.
\eeq
\noindent
Notice that since the elements of $P_{\mu\nu}$ are separable bivectors it has the index structure $P_{\mu\nu\alpha\beta}$.
Explicitly we have

\beq
\label{Plucker2}
P_{\mu\nu\alpha\beta}=Z_{\mu\alpha}Z_{\nu\beta}-Z_{\mu\beta}Z_{\nu\alpha}.
\eeq
\noindent
They are the Pl\"ucker coordinates of six lines in ${\bf CP}^3$ labelled by the subscripts $\mu\nu=01,02,03,12,13,23$.
Alternatively one can introduce the "bivector-valued" vector ${\bf P}$ with six components $P_I$,
$I=01,02,03,12,13,23$ with the same geometric meaning.

\beq
P_I=({\bf A}\wedge{\bf B},{\bf A}\wedge{\bf C},{\bf A}\wedge{\bf D}, {\bf B}\wedge{\bf C},{\bf B}\wedge{\bf D}, {\bf C}\wedge{\bf D})^T.
\eeq
\noindent
In this notation the invariant $I_2$ of Eq. (12) can be written in the form

\beq
I_2=\frac{1}{12}G^{IJ}P_I\cdot P_J=\frac{1}{6}P^{\mu\nu}\cdot P_{\mu\nu}=\frac{1}{6}P^{\mu\nu\alpha\beta}P_{\mu\nu\alpha\beta}.
\eeq
\noindent
Notice that the elements of the Pl\"ucker matrix being now separable bivectors, are multiplied together according to the (\ref{cdot}) law bringing in the $\cdot$ product.
Since $I_2$ contains contractions with
respect to four $SL(2,{\bf C})\otimes SL(2,{\bf C})$ invariant matrices $g$
(two of them operates on the first two and the other two on the last two qubits) it is automatically an $SL(2, {\bf C})^{\otimes 4}$ invariant.

From the form of the Pl\"ucker matrix Eq. (20) it is also clear 
that it encapsulates information concerning  four qubit entanglement in terms of {\it six planes} in ${\bf C}^4$ or alternatively {\it six lines} in ${\bf C}P^3$.
Hence we managed to identify the SLOCC invariant $I_2$ as a                    {\it line invariant} in complex projective space.

Let us now clarify the relationship of our invariant $I_2$ with the ones of Ref. [5].
There also the invariants $L={\rm Det}{\cal L}$, $M={\rm Det}{\cal M}$ and $N={\rm Det}{\cal N}$ were defined where the $4\times 4$ matrices are given by Eqs. (2-4). Notice that our convention for $M$ differs in a sign from the one adopted in Ref. [5]. It can be shown that $M=L+N$ as can be verified by calculating the determinants of the matrices in Eqs (3) and (4) containing  $2 \times 2$ blocks.
Now a straightforward calculation shows that in terms of the algebraically independent invariants $H=2I_1$, $L=I_4$ and $M$ preferred by ref. [5] we have the relation

\beq
6I_2=H^2+2L-4M.
\eeq
\noindent

\subsection{The invariant $I_3$}

In order to understand the geometric meaning of the invariant $I_3$ of Eq. (13)
we have to recall some results from projective geometry.
A {\it plane} in ${\bf CP}^3$ consists of a set of points with homogeneous coordinates $X^{\alpha}$ $\alpha =0,1,2,3$ which satisfy a single linear equation of the form

\beq 
a_{\alpha}X^{\alpha}=0
\eeq
\noindent
where the complex numbers $a_{\alpha}$ are called the {\it coordinates of the plane}. Clearly $a_{\alpha}$ and $\lambda a_{\alpha}$ with $\lambda\neq 0$
determine the same plane in ${\bf CP}^3$ so the set of planes in ${\bf CP}^3$
is  itself a ${\bf CP}^3$ called the {\it dual projective space}.
A plane in ${\bf CP}^3$ is a ${\bf CP}^2$.
There is a unique plane containing $3$ general points in ${\bf CP}^3$. If 
$B_{\beta}$, $C_{\gamma}$ and ${D}_{\delta}$ are $3$ general points then there is a unique solution up to proportionality of the three equations

\beq
a_{\alpha}B^{\alpha}=a_{\alpha}C^{\alpha}=a_{\alpha}D^{\alpha}=0
\eeq
\noindent
given by the first of Eq. (9).
Now we see that the four-vectors $a_{\alpha}$, $b_{\beta}$, $c_{\gamma}$
and $d_{\delta}$ defined in Eq. (9) are the (dual Pl\"ucker) coordinates of four planes in ${\bf CP}^3$. They are defined by the three points $({\bf B}, {\bf C}, {\bf D})$,
$({\bf A}, {\bf C}, {\bf D})$, $({\bf A}, {\bf B}, {\bf D})$ and $({\bf A}, {\bf B}, {\bf C})$ respectively.
Alternatively for these planes we can use the Pl\"ucker coordinates (separable trivectors)
${\bf B}\wedge{\bf C}\wedge{\bf D}$,
${\bf A}\wedge{\bf C}\wedge{\bf D}$, ${\bf A}\wedge{\bf B}\wedge{\bf D}$, ${\bf A}\wedge{\bf B}\wedge{\bf C}$
where for example

\beq
({\bf A}\wedge{\bf B}\wedge{\bf C})_{\alpha\beta\gamma}=
A_{\alpha}B_{\beta}C_{\gamma}+A_{\gamma}B_{\alpha}C_{\beta}+A_{\beta}B_{\gamma}C_{\alpha}-A_{\beta}B_{\alpha}C_{\gamma}-A_{\gamma}B_{\beta}C_{\alpha}-A_{\alpha}
B_{\gamma}C_{\beta}\equiv 3!A_{[\alpha}B_{\beta}C_{\gamma]}.
\eeq
\noindent

The set of points common to two planes is a {\it line} in ${\bf CP}^3$.
A line is given by the points satisfying two linear equations of the form

\beq
a_{\alpha}X^{\alpha}=b_{\alpha}X^{\alpha}=0.
\eeq
\noindent
A sufficient and necessary condition for these equations to hold is

\beq
a_{[\alpha}b_{\beta]}X^{\beta}=0.
\eeq
\noindent
Hence in order to characterize the line (a ${\bf CP}^1$) defined by this equation we can either use the {\it dual Pl\"ucker coordinates} (a separable bivector)

\beq
({\bf a}\wedge{\bf b})_{\alpha\beta}=a_{\alpha}b_{\beta}-a_{\beta}b_{\alpha}\equiv 2!a_{[\alpha}b_{\beta]}
\eeq
\noindent
or using Eq.(9) the {\it Pl\"ucker coordinates} (another separable bivector)

\beq
({\bf C}\wedge{\bf D})_{\gamma\delta}=C_{\gamma}D_{\delta}-C_{\delta}D_{\gamma}\equiv 2!C_{[\gamma}D_{\delta]}.
\eeq
\noindent
This example shows that e.g. the planes with the Pl\"ucker coordinates ${\bf B}\wedge{\bf C}\wedge{\bf D}$ and ${\bf A}\wedge{\bf C}\wedge{\bf D}$ intersect in the projective line given by the Pl\"ucker coordinates ${\bf C}\wedge {\bf D}$.

Now clearly $I_3$ of Eq. (13) is an $SL(2, {\bf C})^{\otimes 4}$ invariant.
Indeed, the dual Pl\"ucker coordinates ${\bf a}$, ${\bf b}$, ${\bf c}$ and ${\bf d}$ are transforming as the $11$, $01$, $10$ and $00$ components of a tensor under transformations of the form $S_1\otimes S_2\otimes I\otimes I$, hence the combination ${\bf a\cdot d}-{\bf b\cdot c}$ is an invariant with respect to such transformations due to the "determinant-like" structure.
Moreover this quantity is also invariant under transformations of the form $I\otimes I\otimes S_3\otimes S_4$ due to the occurrence of the $SL(2, {\bf C})^{\otimes 2}$ invariant $\cdot$ product of Eq.(6).
Notice that $I_3$ has the same structure as $I_1$.
This exemplifies a general pattern: suppose we have an invariant ($I_1$)
, then find a set of {\it covariants} (i.e. ${\bf a}$, {\bf b}$, {\bf c}$ and ${\bf d}$) to construct a new invariant ($I_3$) by exploiting the existing structure of the original invariant ($I_1$). 
It is clear that $I_1$ describes constellations of {\it points} and $I_3$ describes {\it planes} in ${\bf CP}^3$.
It is important to realize however, that unlike $I_3$ the invariant $I_1$ is also a permutation invariant.

Let us also express our invariant in terms of the Pl\"ucker coordinates ${\bf A} \wedge{\bf B}\wedge{\bf C}$ etc as

\beq
I_3=\frac{1}{12}\left(({\bf A}\wedge{\bf C}\wedge{\bf D})\cdot({\bf A}\wedge{\bf B}\wedge{\bf D})-
		({\bf B}\wedge{\bf C}\wedge{\bf D})\cdot({\bf A}\wedge{\bf B}\wedge{\bf C})      \right).
\eeq
\noindent
Alternatively like in Eq. (20) one can define a third order totally antisymmetric Pl\"ucker tensor
$P_{\mu\nu\rho}$ with trivectors as elements.
The  four independent elements of $P_{\mu\nu\rho}$ are
$P_{012}={\bf A}\wedge{\bf B}\wedge{\bf C}$,
$P_{123}={\bf B}\wedge{\bf C}\wedge{\bf D}$,
$P_{023}={\bf A}\wedge{\bf C}\wedge{\bf D}$ and
$P_{013}={\bf A}\wedge{\bf B}\wedge{\bf D}$.
Notice that these quantities have the index structure $P_{012\alpha\beta\gamma}=({\bf A}\wedge{\bf B}\wedge{\bf C})_{\alpha\beta\gamma}$ where the definition of Eq. (27) holds.
Using this notation we
have

\beq
I_3=\frac{1}{12^2}P^{\mu\nu\rho}\cdot P_{\mu\nu\rho}=\frac{1}{12^2}P^{\mu\nu\rho\alpha\beta\gamma}P_{\mu\nu\rho\alpha\beta\gamma},
\eeq
\noindent
an expression to be compared with Eq. (23) obtained for our line invariant.

Finally let us relate our invariant $I_3$ to the sixth order ones of Ref. [5].
Define the quadrilinear form

\beq
Z({\bf x},{\bf y},{\bf z},{\bf t})=\sum_{i,j,k,l=0}^1Z_{ijkl}x_iy_jz_kt_l.
\eeq
\noindent
Using this for each pair of variables one then defines the covariants (now we define it for the pair $xy$)

\beq
b_{xy}({\bf x}, {\bf z})={\rm Det}\left(\frac{{\partial}^2Z}{{\partial}z_i{\partial}t_j}\right).
\eeq
\noindent
Now let us reinterpret these biquadratic forms as bilinear forms on $S^2{\bf C}$ (the symmetric part of ${\bf C}^2\otimes {\bf C}^2$) as

\beq
b_{xy}({\bf x}, {\bf y})=(x_0^2,x_0x_1,x_1^2)B_{xy}\begin{pmatrix} y_0^2\\y_0y_1\\y_1^2\end{pmatrix},
\eeq
\noindent
i.e. $B_{xy}$ is a $3\times 3$ matrix.
Then following Ref. [5] we define

\beq
D_{uv}={\rm Det}(B_{uv}).
\eeq
\noindent
Hence we have six sextic invariants $D_{xy}$, $D_{zt}$, $D_{xz}$, $D_{yt}$,
$D_{xt}$ and $D_{yz}$.
According to Ref. [5] only four of them is independent due to the relations
$D_{xy}=D_{zt}$, $D_{xz}=D_{yt}$ and $D_{xt}=D_{yz}$.
Now a straightforward calculation shows that

\beq
I_3=\frac{1}{2}(D_{xz}+D_{xt}).
\eeq
\noindent
In Ref. [5] the authors used the invariant $D\equiv D_{xt}$ as a fundamental one satisfying the relation
$D_{xz}-D_{xt}=HL$
hence we can write
\beq
I_3=D+\frac{1}{2}HL.
\eeq
\noindent
By virtue of Eqs. (14), (16), (24) and (39) the relationship between
our set of invariants $(I_1,I_2,I_3,I_4)$ and the ones $(H,M,L,D)$ used in Ref. [5] is established.

\subsection{The invariants $I_4$ and $I_2$}

The meaning of this invariant is clear.
$L={\rm Det}{\cal L}$ is vanishing when the vectors ${\bf A}$, ${\bf B}$, ${\bf C}$ and ${\bf D}$ are linearly dependent. 
Since $I_4$ and $I_2$ are both of fourth order let us now explore the relationship between them.

Let us label the six lines as in Eq. (20). Hence for example $P_{01}$ is the line corresponding to the separable bivector ${\bf A}\wedge{\bf B}$. More precisely this object has the index structure $(P_{01})_{\alpha\beta}=({\bf A}\wedge{\bf B})_{\alpha\beta}=A_{\alpha}B_{\beta}-A_{\beta}B_{\alpha}$. 
In this notation

\beq
L=\frac{1}{4}{\varepsilon}^{\alpha\beta\gamma\delta}(P_{01})_{\alpha\beta}(P_{23})_{\gamma \delta}.
\eeq
\noindent
We can regard this expression as a symmetric bilinear form in the six Pl\"ucker coordinates of the two lines ${\bf A}\wedge{\bf B}$ and ${\bf C}\wedge{\bf D}$.
Let us denote this bilinear form by $\langle ,\rangle$ hence we have

\beq
\langle ,\rangle:\bigwedge{\bf C}^4\otimes {\bigwedge}{\bf C}^4\to {\bf C},\qquad
(P_{\mu\nu},P_{\rho\sigma})\mapsto \langle P_{\mu\nu}, P_{\rho\sigma}\rangle. 
\eeq
\noindent
Defining the dual of a bivector as
\beq
{^{\ast}P_{\alpha\beta}}=\frac{1}{2}{\varepsilon}_{\alpha\beta\gamma\delta}P^{\gamma\delta},
\eeq
\noindent
it is easy to show that
\beq
\langle P_{\mu\nu}, P_{\rho\sigma}\rangle=\langle{^{\ast}P_{\mu\nu}}, {^{\ast}P_{\rho\sigma}}\rangle.
\eeq
\noindent
In this notation the equation $\langle P_{\mu\nu}, P_{\nu\rho}\rangle=0$ expresses the fact that the planes described by the separable bivectors $P_{\mu\nu}$ and $P_{\nu\rho}$ in ${\bf C}^4$ have a line in common. In the ${\bf CP}^3$ picture this is equivalent to the fact that the corresponding {\it lines} in ${\bf CP}^3$ intersect {\it in a point}. Hence we can look at $L$ as a line invariant too, moreover we have
the obvious relations

\beq
I_4=\langle P_{01},P_{23}\rangle=
\langle P_{02},P_{31}\rangle=
\langle P_{03},P_{12}\rangle=
\langle {^{\ast}P_{01}},{^{\ast}P_{23}}\rangle=
\langle {^{\ast}P_{02}},{^{\ast}P_{31}}\rangle=
\langle {^{\ast}P_{03}},{^{\ast}P_{12}}\rangle.
\eeq
\noindent

Let us now look at quantities like $\langle P_{\mu\nu},{^{\ast}P_{\rho\sigma}}\rangle$!  
We have for example

\beq
\langle P_{01}, {^{\ast}P_{23}}\rangle=({\bf A}\cdot{\bf C})({\bf B}\cdot{\bf D})-
                         ({\bf A}\cdot{\bf D})({\bf B}\cdot{\bf C}=\frac{1}{2}({\bf A}\wedge{\bf B})\cdot ({\bf C}\wedge{\bf D}).
\eeq
\noindent
Since $\langle P_{\mu\nu},{^{\ast}P_{\rho\sigma}}\rangle=\langle{^{\ast}P_{\mu\nu}},P_{\rho\sigma}\rangle$ one can write now the invariant $6I_2$ in the form

\beq
6I_2=\langle P_{01},{^{\ast}P_{23}}\rangle+\langle {^{\ast}P_{01},P_{23}}\rangle)+
\langle P_{02},{^{\ast}P_{13}}\rangle +\langle {^{\ast}P_{02}},P_{13}\rangle -\langle P_{03},{^{\ast}P_{03}}\rangle -\langle P_{12},{^{\ast}P_{12}}\rangle .
\eeq
\noindent
Using Eq. (43) let us calculate $6(I_1\pm I_4)$! 
We get

\beq
6(I_2\pm I_4)=\pm\langle P_{01}\pm {^{\ast}P_{01}},P_{23}\pm {^{\ast} P_{23}}\rangle
\pm\langle P_{02}\mp {^{\ast}P_{02}},P_{31}\mp {^{\ast} P_{31}}\rangle
-\langle P_{12}\mp{^{\ast}P_{03}},{^{\ast}P_{12}}\mp P_{03}\rangle.
\eeq
\noindent
As we will see these invariants will occur in the expression for the hyperdeterminant. The geometric meaning of these invariants is connected to the intersection properties of the {\it self-dual} (${^{\ast}P}= P$) or {\it anti-self-dual} (${^{\ast}P}=-P$) parts of the planes in ${\bf C}^4$ (or alternatively of lines in ${\bf CP}^3$).
For example for $P_{01}$ self-dual, $P_{31}$ anti-self-dual and $P_{12}$ identical to the dual line of $P_{30}$ (an equivalent condition for this is $L({\bf B}\wedge{\bf C})=-{\bf b}\wedge{\bf c}$) the invariant $I_4-I_2$ vanishes.
It is easy to check that the invariant $U$ occurring in Ref. [5] can be related to one of these invariants as
\beq
U\equiv H^2-4(L+M)=6(I_4-I_2).
\eeq
\noindent
The fact that (among others) this invariant might have a geometric meaning was raised in Ref.[5].

\section{The hyperdeterminant}

Let us now consider the hyperdeterminant $D_4$ for the four-qubit system.
As it is well-known for two-qubit systems the determinant $D_2=Z_{00}Z_{11}-Z_{01}Z_{10}$
is related to the {\it concurrence}\cite{Kundu} as ${\cal C}=2\vert D_2\vert $ characterizing two-qubit entanglement. Similarly for three-qubits the basic quantity is the three-tangle\cite{Kundu} 
${\tau}=4\vert D_3\vert$ which is related to the hyperdeterminant $D_3$ of a $2\times 2\times 2$ tensor formed from the $8$ complex amplitudes $Z_{ijk}$.
$D_3$ is an irreducible polynomial in the $8$ amplitudes which is the sum of $12$ terms of degree four.
For the explicit expression of $D_3$ see e.g. the book of Gelfand et.al.\cite{Gelfand}.
It is known that the next item in the line namely the hyperdeterminant $D_4$ of format $2\times 2\times 2\times 2$ is a polynomial of degree $24$ in the $16$ amplitudes $Z_{ijkl}$ which has 2894276 terms\cite{Grier}.
An expression in terms of the fundamental invariants $(H,L,M,D)$ was given
in Ref. [5].
Here we are interested in the explicit form of $D_4$ the hyperdeterminant of the $2\times 2\times 2\times 2$ tensor $Z_{ijkl}$ based on the special invariants  $(I_1,I_2,I_3,I_4)$  we have found in our ${\bf CP}^3$ picture.

As it is well known\cite{Gelfand} the hyperdeterminant $D_4$ is the unique irreducible polynomial in the $16$ unknowns $Z_{ijkl}$ that vanishes whenever the system of equations 

\beq
F=\frac{\partial F}{\partial x}=
\frac{\partial F}{\partial y}=\frac{\partial F}{\partial z}=\frac{\partial F}{\partial t}=0,
\eeq
\noindent
where
\beq 
F=Z_{0000}+Z_{0001}t+Z_{0010}z+Z_{0100}y+Z_{1000}x+Z_{0011}zt+Z_{0101}yt+\dots
+Z_{1110}xyz+Z_{1111}xyzt
\eeq
\noindent
has a solution $(x_0,y_0,z_0,t_0)$ in ${\bf C}^4$.

Using the method of Schl\"afli according to Theorem 14.4.1 and Corollary 14.2.10 of Ref. [20]
$D_4$ coincides with the discriminant ${\Delta}$ of $D_3(Z_{0jkl}+{\lambda}Z_{1jkl})$ considered as a polynomial in ${\lambda}$ divided by $256$.
This method has already been used to obtain a much simpler form for $D_3$ of geometric meaning\cite{Lev1}.
For $D_4$ a method equivalent to this has been applied with the result\cite{Luque}

\beq
256D_4=S^3-27T^2
\eeq
\noindent
where
\beq
12S=U^2-2V,\quad 216T=U^3-3UV+216D^2,
\eeq
\noindent
with
\beq
U=H^2-4(L+M),\quad V=12(HD+2LM).
\eeq
\noindent

Let us now express $D_4$ in terms of our invariants $(I_1,I_2,I_3,I_4)$ of geometric significance. Using relations (14), (16), (24) and (39) in Eqs. (51-53) we obtain the result

\beq
S=(I_4^2-I_2^2)+4(I_2^2-I_1I_3),\quad T=(I_4^2-I_2^2)(I_1^2-I_2)+(I_3-I_1I_2)^2.
\eeq
\noindent
In this form it is obvious that the combined invariants
$I_4\pm I_2$, $I_1^2-I_2$ of fourth $(I_3-I_1I_2)$ of sixth and $I_2^2-I_1I_3$
of eight order should play a basic geometric role.
We have already clarified the geometric meaning of the first two invariants.
They are related to self-duality and anti-self-duality of the corresponding lines in ${\bf CP}^3$.
One of these invariants $I_4-I_2$ is just $\frac{1}{6}U$ also used in Ref. [5].
In our form of $D_4$ we prefer to also use the dual combination $I_4+I_2$.
For the time being we do not know any geometrical interpretation of the other
combinations.
Intuitively it is clear that the invariant $I_1^2-I_2$ should play a similar role than the other fourth order invariants.
Indeed we have chosen the third and fourth qubits to play a special role.
(An equivalent picture arises when a special role is assigned to the first and the second qubit.)
The ${\bf C}^4$ defined by them is equipped with the bilinear form Eq. (6).
The null vectors (i.e. the ones satisfying ${\bf X}\cdot {\bf X}=0$)
describe a quadric embedded in ${\bf CP}^3$ which is isomorphic to ${\bf CP}^1\times {\bf CP}^1$, i.e. it is ruled by two families of projective lines
which can be shown to be self-dual or anti-self dual respectively\cite{Lev3}.
Projective lines lying entirely inside a fixed quadric are called isotropic lines.
Had we chosen the two qubits playing a special role differently
the notion of self or anti-self-duality of isotropic lines would have been defined with respect to a different quadric. In this picture we conjecture that the invariants $I_4-I_2$ and $I_1^2-I_2$ would play a dual role. 
Since altogether we have three inequivalent choices then we can conclude that the fourth order invariants are related to the notion of duality of isotropic lines with respect to a fixed quadric in ${\bf CP}^3$.
It would be interesting to find a geometric interpretation for the remaining invariants too.

Let us now consider another interesting property of $D_4$ expressed in terms of our invariants $(I_1, I_2, I_3, I_4)$.
As it is well-known the discriminant ${\Delta}$ of the polynomial
$e_4w^4+e_3w^3+e_2w^2+e_1w+e_0$
is given by the expression (see Ref. [20] Eq. (1.35) on page 405)
\begin{eqnarray}
\Delta(e_4w^4&+&e_3w^3+e_2w^2+e_1w+e_0)=256e_0^3e_4^2-192e_0^2e_1e_3e_4^2-128e_0^2e_2^2e_4^2\nonumber\\&+&144e_0^2e_2e_3^2e_4-27e_0^2e_3^4+144e_0e_1^2e_2e_4^2
-6e_0e_1^2e_3^2e_4-80e_0e_1e_2^2e_3e_4\nonumber\\&+&18e_0e_1e_2e_3^3+16e_0e_2^4e_4-4e_0e_2^3e_3^3-27e_1^4e_4^2+18e_1^3e_2e_3e_4-4e_1^3e_3^3\nonumber\\&-&
4e_1^2e_2^3e_4+e_1^2e_2^2e_3^4.
\end{eqnarray}
\noindent
Let us now consider the polynomial of the special form
 
\beq
p[I_1,I_2,I_3,I_4; w]\equiv w^4-(4I_1)w^3+(6I_2)w^2-(4I_3)w+I_4^2
\eeq
\noindent
Then a straightforward calculation shows that

\beq
256D_4=\Delta( p[I_1,I_2,I_3,I_4; w]).
\eeq
\noindent

Notice that the polynomial $p$ is not directly related to the one arising from the method of Schl\"afli.
In this case one obtains
\beq
D_3(Z_{0jkl}+{\lambda}Z_{1jkl})=h_4{\lambda}^4+h_3{\lambda}^3+h_2{\lambda}^2+h_1{\lambda}+h_0
\eeq
\noindent
where unlike the ones $e_s$ the coefficients $h_s\quad s=0,1\dots 4$ are fourth order polynomials of the $Z_{ijkl}$ that are {\it not invariant} with respect to the full group $SL(2, {\bf C})^{\otimes 4}$.
However, the discriminant of this polynomial again gives $256D_4$ which {\it is}
already an invariant with respect to the full group of SLOCC transformations.

In order to illustrate the advantages of using our invariants $(I_1,I_2,I_3,I_4)$ let us now calculate their values on the generic SLOCC class.
A generic pure state of four qubits can always be transformed to the form\cite{Verstraete}

\begin{eqnarray}
\vert G_{abcd}\rangle&=&\frac{a+d}{2}(\vert 0000\rangle +\vert 1111\rangle)
+\frac{a-d}{2}(\vert 0011\rangle +\vert 1100\rangle)\nonumber\\&+&
\frac{b+c}{2}(\vert 0101\rangle +\vert 1010\rangle)+
\frac{b-c}{2}(\vert 0110\rangle +\vert 0110\rangle),
\end{eqnarray}
\noindent
where $a,b,c,d$ are complex numbers.
For this state the reduced density matrices obtained by tracing out all but one party are proportional to the identity. This is the state with maximal four-partite entanglement.
Another interesting property of this state is that it does not contain true three-partite entanglement\cite{Verstraete}.
A straightforward calculation shows that the values of our invariants $(I_1,I_2,I_3,I_4)$ occurring for the state $\vert G_{abcd}\rangle$ representing the generic SLOCC class are

\beq
I_1=\frac{1}{4}[a^2+b^2+c^2+d^2],\quad
I_2=\frac{1}{6}[(ab)^2+(ac)^2+(ad)^2+(bc)^2+(bd)^2+(cd)^2],
\eeq
\noindent
\beq
I_3=\frac{1}{4}[(abc)^2+(abd)^2+(acd)^2+(bcd)^2],\quad I_4=abcd,
\eeq
\noindent
hence the values of the invariants $(4I_1,6I_2,4I_3,I_4^2)$ occurring in the polynomial Eq.(56) are given in terms of the elementary symmetric polynomials in the
variables $(x_1,x_2,x_3,x_4)=(a^2,b^2,c^2,d^2)$.
From this and Eq. (57) it immediately follows that the value of the hyperdeterminant on the SLOCC orbit represented by the state $\vert G_{abcd}\rangle$ is

\beq
D_4=\frac{1}{256}\Pi_{i<j}(x_i-x_j)^2=\frac{1}{256}V(a^2,b^2,c^2,d^2)^2,
\eeq
\noindent
in accordance with Ref.[5] where $V$ is the Vandermonde determinant.
For the other SLOCC classes and the values of the invariants $(H,L,M,D)$
see Ref. [5]. It is straightforward to give the alternative values of $(I_1,I_2,I_3,I_4)$ on these classes.

\section {Comments and conclusions}

In this paper we have considered some aspects of the problem of understanding four qubit entanglement in geometric terms.
We have replaced two from the set containing four algebraically independent invariants of Ref. [5] by new ones. In this way all four invariants have a simple geometric meaning. $I_1$ is based on $4$ 0-planes (points), $I_2$ on $6$ 1-planes (lines), $I_3$ on $4$ 2-planes (planes) and finally $I_4$ on a single $3$-plane
in ${\bf CP}^3$.
According to Theorem 2. of Ref.[24] the magnitudes of these invariants can be used as entanglement monotones characterizing four-qubit entanglement. 
Moreover, for an arbitrary four-qubit state after calculating the set of invariants $(I_1,I_2,I_3,I_4)$ and the value of the hyperdeterminant $D_4$ , for $D_4\neq 0$ we obtain four different roots of the fourth order equation Eq. (56).
These roots are just the complex numbers $(a^2,b^2,c^2,d^2)$.
Their square roots produce the values $(\pm a,\pm b,\pm c,\pm d)$
appearing in the canonical form $G_{abcd}$.
This shows that the study of the degenerate cases of multiple roots of Eq. (56) arising for $D_4=0$, could be useful for obtaining the parameters of the canonical forms\cite{Verstraete} from the values of the basic invariants.
This process is similar to the spirit of the one found for the three-qubit case\cite{Acin}. There by calculating the values of the independent $SU(2)^{\otimes 3}$ invariants
the canonical form of an arbitrary three-qubit state was found.

We note that our four qubit entanglement monotones fit nicely into the scheme of Ref. [9] generating a class of $n$-qubit entangled monotones based on bipartite decompositions of ${\cal H}={\bf C}^{2^n}$. 
The basic idea followed there was to consider the manifold of subspaces of ${\cal H}$ i.e. suitable Grassmannians with the corresponding Pl\"ucker coordinates for them.
Equivalently we should consider subspaces of the corresponding projective spaces
${\bf P}({\cal H})$.
Fixing a quadric ${\cal Q}$ defined by a bilinear form similar to Eq. (6)
a class of SLOCC invariants expressed in terms of these Pl\"ucker coordinates can be generated.
In this way we were able to reproduce three of the basic four-qubit invariants
(i.e. the triple $(H,L,M)$).
Now we see that by employing also the notion of projective duality all four algebraically independent invariants of the four-qubit case can be written in the Pl\"ucker form. (See Eqs. (33) and (23) for our new invariants).

Finally let us propose a suggestive geometric picture for four-qubit entanglement.
In the usual picture\cite{Miyake} a four-qubit state can be represented
by a {\it single point} in ${\bf CP}^{15}$.
Different SLOCC classes correspond to this point lying on different
subvarieties in ${\bf CP}^{15}$.
Here we would like to suggest an alternative picture.
To a four-qubit state we associate a set of {\it four points} ${\bf A}$, ${\bf B}$,   ${\bf C}$, ${\bf D}$,  {\it six lines} ${\bf A}\wedge{\bf B}$, ${\bf A}\wedge{\bf C}$, ${\bf A}\wedge{\bf D}$, ${\bf B}\wedge{\bf C}$, ${\bf B}\wedge {\bf D}$,       ${\bf C}\wedge{\bf D}$ and {\it four planes} ${\bf A}\wedge{\bf B}\wedge{\bf C}$,    ${\bf A}\wedge{\bf C}\wedge{\bf D}$, ${\bf A}\wedge{\bf B}\wedge{\bf D}$,        ${\bf B}\wedge{\bf C}\wedge{\bf D}$ in the space ${\bf CP}^3$ of smaller dimension. It is easy to see looking at the intersection properties of these geometrical objects that they are forming a {\it tetrahedron} in ${\bf CP}^3$. 
This correspondence between entangled states and geometric objects (unlike the previous one) is {\it nonlocal}.
The invariants $(I_1,I_2,I_3,I_4)$ we have proposed obviously characterize
the properties of this terahedron. For example for $I_4=0$ the four points corresponding to the four vectors ${\bf A}$, ${\bf B}$, ${\bf C}$ and ${\bf D}$  in ${\bf C}^4$ are not linearly independent (some of them are proportional), hence the tetrahedron is degenerating
to a triangle or a line etc. depending on the degree of degeneracy.
We conjecture that the vanishing of the other three invariants somehow characterize more intricate degeneracies occurring with lines and planes of the tetrahedron.
The class containing no degeneracy is the $G_{abcd}$-class that can be represented by a regular terahedron.
It would be interesting to understand how the SLOCC classes arise in this picture.

Let us elucidate the meaning of the proposed correspondence a little bit further.
The usual geometric classification schemes
for multiqubit systems are based on the use of hyperdeterminants of more general type. These hyperdeterminants describe geometrically hypersurfaces
projectively dual to the so called Segre embedding\cite{Gelfand,Miyake} representing the subvariety of totally separable states\cite{Heydari}. 
For two, three and four-qubits for instance the manifold of totally separable states is ${\bf CP}^1\times {\bf CP}^1$ embedded in ${\bf CP}^3$ ,${\bf CP}^1\times {\bf CP}^1\times {\bf CP}^1$ in ${\bf CP}^7$ ,
and ${\bf CP}^1\times {\bf CP}^1\times {\bf CP}^1\times {\bf CP}^1$ in ${\bf CP}^{15}$ respectively. Here $n=2,3,4$ qubit states carrying entanglement are represented by {\it points} off the Segre surfaces. Such surfaces are representing totally separable states in {\it different} projective spaces (${\bf CP}^{2^n-1}$). 
Here following the spirit of our previous set of papers\cite{Levay,Lev1,Lev2,Lev3}
we prefer to suggest a unified  ${\bf CP}^3$ picture.
For $n=2,3,4$ we take ${\bf CP}^3$ with a fixed quadric ${\cal Q}$ based on our choice of bilinear form Eq. (6). Points lying on ${\cal Q}$ as a subvariety of ${\bf CP}^3$ correspond to {\it null vectors} in ${\bf C}^4$.
For $n=2$ to a separable or an entangled state corresponds a point {\it on}  or {\it off} ${\cal Q}$
respectively.
For $n=3$ we get the following geometric picture\cite{Lev1,Lev3}. To a three-qubit state in the $GHZ$-class  
corresponds a {\it line} in ${\bf CP}^3$ intersecting ${\cal Q}$ at {\it two} points. To a state in the $W$-class corresponds a line tangent to ${\cal Q}$ at a {\it point}.
The separable classes $B(AC)$ and $C(AB)$ are represented by isotropic lines 
lying entirely in ${\cal Q}$. They are self-dual and anti-self-dual lines
belonging to the two different rulings of ${\cal Q}$.
The $A(BC)$ and $(A)(B)(C)$ classes again correspond to the degenerate case of
points {\it on} and {\it off} the quadric ${\cal Q}$ (see also the pictorial representation of Ref.[22]).
We expect a similar pattern to exist also for the four-qubit ($n=4$) case.
Here we have more lines arranged to form a tetrahedron, and we have to consider
constellations of these lines with respect to our fixed quadric ${\cal Q}$.
The picture arising in this way has some striking similarity with the Majorana\cite{Majorana} representation of states with spin $s$.
One can represent geometrically a state of spin $s$ as a {\it single point} in ${\bf CP}^{2s}$, or alternatively as a {\it constellation of $2s$ points}
on ${\bf CP}^1$ i.e. the Bloch sphere.
Some degeneracies can occur in this case e.g. when $2s$ points degenerate to a single one with multiplicity $2s$ corresponding to the states of highest and lowest weights.
In the same spirit we would rather represent $n$-qubit entangled states in ${\bf CP}^3$. Here, however in order to account for the nonlocality of multiqubit quantum entanglement instead of merely a collection points we have to also consider constellations of lines and planes in ${\bf CP}^3$.
Though this analogy is very appealing we expect it to run out of steam for the $n=5$ case (five-qubits) where probably we should furnish ${\bf CP}^3$ with more extra structures than a quadric.
However for $n \geq 5$ in principle we can consider constellations of simplexes in ${\bf CP}^3$ related to fundamental invariants of the SLOCC group
whose combinatorial variability should somehow correspond to the proliferating number of entanglement classes.

\section{Acknowledgements}
Financial support from the Orsz\'agos Tudom\'anyos Kutat\'asi Alap
(grant numbers T047035, T047041, T038191) is
gratefully acknowledged.

\end{document}